\documentclass[final,5p,times,twocolumn]{elsarticle}

\usepackage{color}

\journal{Computational Materials Science}

\begin{document}

\begin{frontmatter}

\title{Electric field effect in boron and nitrogen doped graphene bilayers}

\author[1,2]{G. A. Nemnes\corref{mycorrespondingauthor}}
\cortext[mycorrespondingauthor]{Corresponding author. Tel.: +40 (0)21 457 4949/157. \\ {\it E-mail address:} nemnes@solid.fizica.unibuc.ro (G.A. Nemnes).}
\address[1]{University of Bucharest, Faculty of Physics, Materials and Devices for Electronics and Optoelectronics Research Center, 077125 Magurele-Ilfov, Romania}
\address[2]{Horia Hulubei National Institute for Physics and Nuclear Engineering, 077126 Magurele-Ilfov, Romania}

\author[2]{T. L. Mitran}

\author[3]{A. Manolescu}
\address[3]{School of Science and Engineering,
        Reykjavik University, Menntavegur 1, IS-101 Reykjavik, Iceland}

\author[1,4]{Daniela Dragoman}
\address[4]{Academy of Romanian Scientists, Splaiul Independentei 54, Bucharest 050094, Romania}

\begin{abstract}
Unlike single layer graphene, in the case of $AB$-stacked bilayer graphene
(BLG) one can induce a non-zero energy gap by breaking the inversion symmetry 
between the two layers using a perpendicular electric field. This is an essential 
requirement in field-effect applications, particularly since the induced gap in BLG systems can be 
further tuned by the magnitude of the external electric field. Doping is another way to modify the electronic
properties of graphene based systems. We investigate here BLG systems doped with boron and nitrogen in the
presence of external electric field, in the framework of density functional
theory (DFT) calculations. Highly doped BLG systems are known to behave
as degenerate semiconductors, where the Fermi energy depends on the
doping concentration but, in addition, we show that the electronic properties drastically depend also on the applied electric field. By changing the magnitude and the orientation of the electric field, the gap size and position relative to the Fermi 
level may be tuned, essentially controlling the effect of the extrinsic doping. In this context, 
we discuss in how far the external electric field may suitably adjust the effective doping and, 
implicitly, the conduction properties of doped BLG systems.
\end{abstract}


\end{frontmatter}


\section{Introduction}

With the ongoing progress in the development of single layer graphene (SLG) \cite{Geim2007} based electronic devices, few layer graphene (FLG) \cite{PARK20101088} receives more and more attention. In particular, bilayer graphene (BLG) shares many of the electronic, mechanical and chemical properties with SLG, however with a notable difference. While SLG suffers from the well know gap problem, in bilayer graphene one can induce a tunable electronic gap by the use of an external electric field \cite{PhysRevB.74.161403,Zhang2009}. This is significant as it opens the path for a wide variety of field effect applications like field-effect transistors (FETs) with high on/off current ratio \cite{Xia2010},
thermoelectric devices \cite{Chien2016}, 
graphene based bolometers as photo-detectors \cite{Yan2012},
electrically controlled adsorption of oxygen \cite{sato2011},
electronic spin transport \cite{Avsar2016} and
modulation of spin-orbit interaction for spin FETs \cite{Afzal2018}, as well as applications in photonics \cite{Hugen2015}. 

Engineering a band gap in SLG is problematic as the different procedures involved, e.g. by limiting the lateral width in graphene nanoribbons, hybridization induced by adsorbed molecules, chemical modifications or using the influence of substrates, may significantly alter the properties of pristine graphene. Reducing the symmetry from $C_{6v}$ to $C_{3v}$ by breaking the symmetry of $A$ and $B$ sublattices is employed in most cases, being a generic method for inducing a band gap in SLG \cite{C4MH00124A}. In the pristine BLG system with Bernal $AB$ stacking, the inversion symmetry breaking between the two layers can be achieved by applying an external electric field, leading to the appearance of a tunable gap, predicted theoretically \cite{PhysRevB.74.161403,0034-4885-76-5-056503} and observed experimentally \cite{Zhang2009,Ohta951,PhysRevLett.99.216802,PhysRevLett.102.256405}. This is of practical interest, since similarly to the silicon based standard CMOS technology, one may envision integrated circuits fabricated on single BLGs, with extrinsic p-doped and n-doped regions embedded with semiconducting regions obtained by field control exerted by top gates.

Boron and nitrogen doped SLG and BLG were synthesized by arc discharge using boron-packed graphite electrodes and in the presence of ammonia, respectively \cite{doi:10.1002/adma.200901285}.
First-principles electronic band structure calculations of the doped BLG systems consistently explain the n- and p-type behavior and scanning tunneling microscopy (STM) images \cite{PhysRevB.85.035444,FUJIMOTO201557}. Tight-binding models have been also employed in the description of doped BLG systems \cite{PhysRevB.80.165406,MOUSAVI201890}. 
Furthermore, in the absence of an electric field the gap of BLG can be chemically tuned by doping \cite{DENIS2010251,PhysRevB.82.245414,Fujimoto2015} and functionalization \cite{PhysRevB.78.085413,HU201475}. 
Applying strain on doped BLG is another approach, by which the mechanically tunable electronic energy gap can be achieved \cite{doi:10.1021/nl101617x}.

Highly doped BLG systems behave as degenerate semiconductors and are well suited as contact regions and interconnects, which can be more easily embedded in the host semiconductor compared to metallic electrodes. Moreover, degenerate semiconductors are typically employed as transparent conductors, presenting high conductivity even at low temperatures.
We investigate here the combined effect of doping and external electric field in BLG systems, which has not been explored in detail so far, and yet it is crucial in the context of gap tuning by stacked gates. We show that the Fermi energy can be modified by two parameters, one intrinsic and one extrinsic, namely the doping concentration and the electric field. This suggests the possibility of adjusting the conduction properties and allows for a local reconfiguration of device attributes, for example in p-n junctions with tunable chemical potentials \cite{Grover2017}.

\section{Model systems and computational methods}

\begin{figure}[t]
\centering
\includegraphics[width=4.0cm]{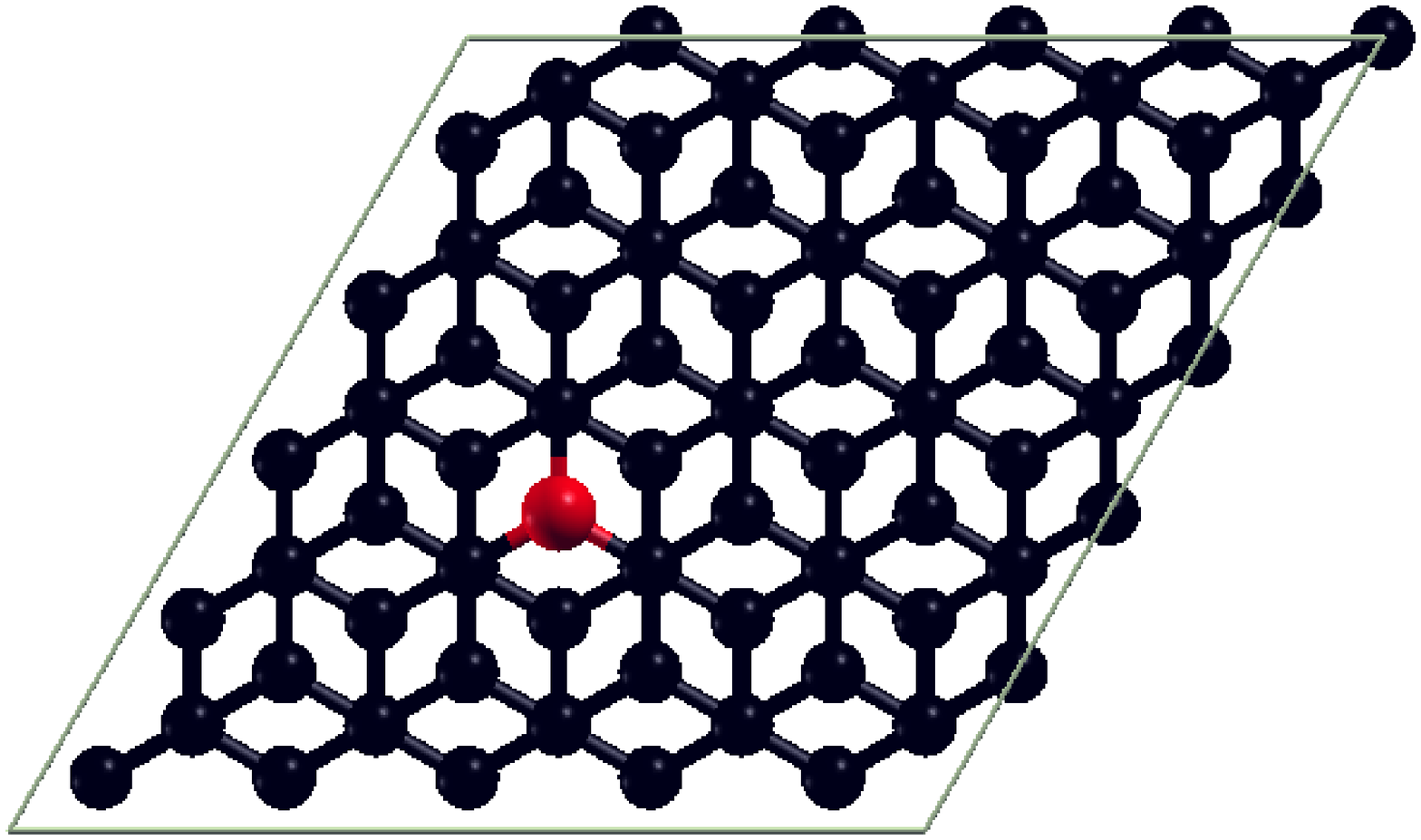}\hspace*{0.1cm}
\includegraphics[width=4.0cm]{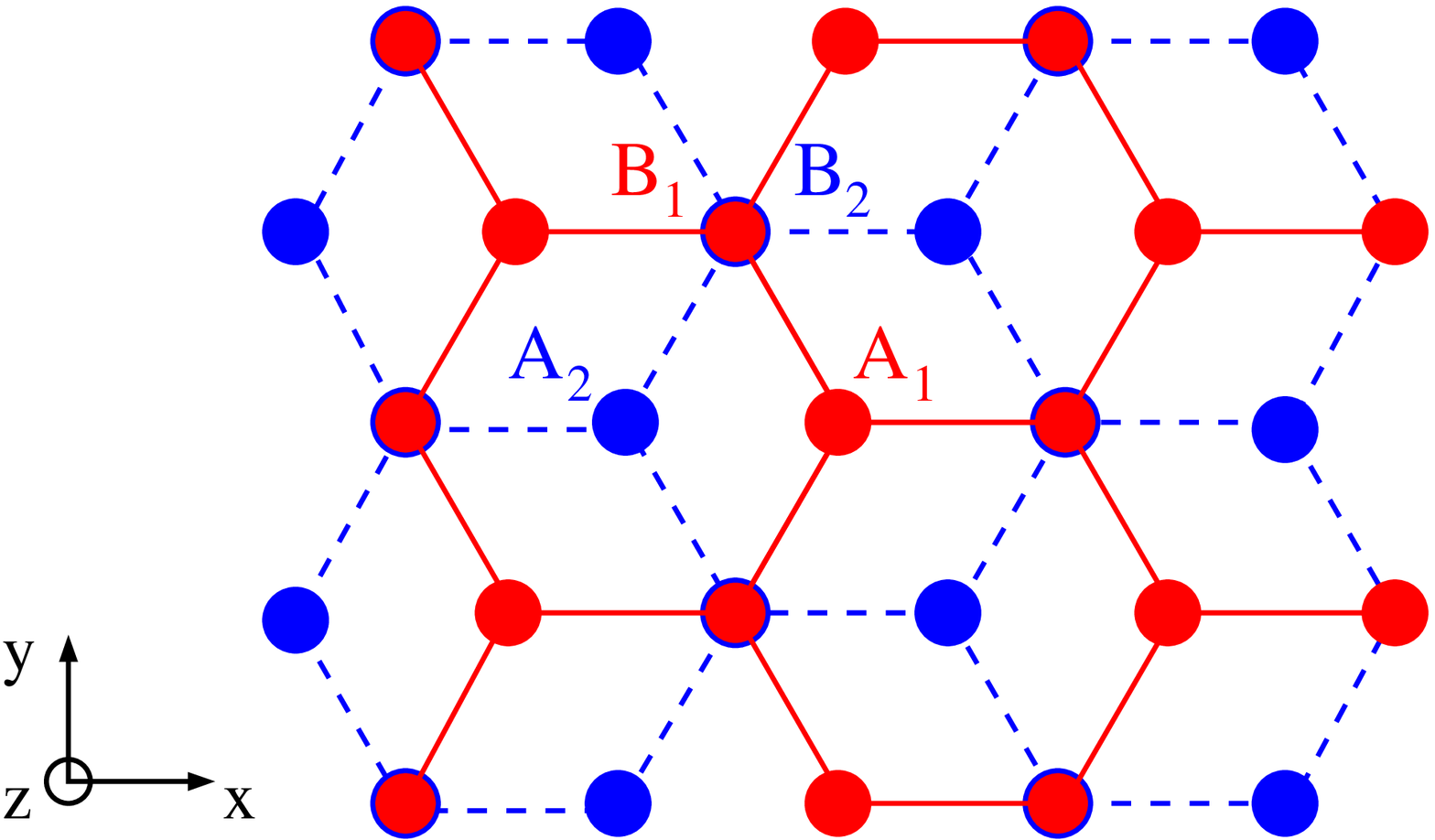}\\
(a)\hspace*{5cm}(b)
\caption{(a) Top view of the $5\times5$ BLG super cell, with one impurity (boron or nitrogen) in the top layer, marked by the red dot. (b) A sketch of the BLG indicating the $A$ and $B$ type positions: $A_1$, $B_1$ in the top layer and $A_2$, $B_2$ in the bottom layer.}
\label{struct}
\end{figure}

We consider boron and nitrogen doped BLG systems with $AB$ stacking, which is the ground state configuration, while $AA$ stacking corresponds to a higher energy state. A $5\times5$ supercell with 100 atoms is depicted in Fig.\ \ref{struct}, showing the $A$ and $B$ type positions in the two layers: in top view $B_1$ and $B_2$ sites overlap, in contrast to $A_1$ and $A_2$ sites. In the following, we analyze the influence of the dopant site for systems with one impurity in the supercell, placed in the top layer and two impurities in the supercell, one in each layer. Larger systems are also investigated in connection with decreasing the doping concentration, up to $14\times14$ supercells containing 784 atoms. Different impurity configurations are also explored in order to analyze in more detail the potential modifications of the electronic properties.

The DFT calculations are performed using the SIESTA code \cite{0953-8984-14-11-302} employing the vdW-DF functional of Dion {\it et al.} (DRSLL) \cite{PhysRevLett.92.246401}, which accurately reproduces the experimental interlayer distance of $3.35$\AA\ and allows a consistent description of the electronic properties \cite{Birowska_Milowska_Majewski_2011}. The self-consistent solution of Kohn-Sham equations was obtained using the DZ basis set, a grid cutoff of 500 Ry and norm-conserving pseudopotentials of Troullier and Martins \cite{PhysRevB.43.1993} with a typical valence electron configurations for carbon, boron and nitrogen.
For the Brillouin-zone integration a $11\times11\times1$ k-points were used in the Monkhorst-Pack scheme, while for generating the density of states (DOS) a finer grid of $101\times101\times1$ k-points is used. In the supercell approach the atomic positions are optimized until the residual forces are less than 0.04 eV/\AA. For the doped BLG systems considered here, the C-B and C-N bond lengths are sharply peaked around 1.47\AA\ and 1.41\AA, respectively. The C-B bond length is considerably larger than the C-C bond length of pristine graphene of 1.42\AA, in contrast to the C-N bond which is slightly shorter than the C-C bond, as reported in other studies \cite{doi:10.1021/acs.jpcc.5b07359,C2RA22664B}. For the systems considered here, the relatively low doping has little influence on the interlayer distance so that it remains practically unchanged. 
The electronic properties of doped BLGs are analyzed considering external electric fields up to a few V/nm, applied perpendicular to the graphene planes, as they are typically employed in order to open the energy gap in pristine BLGs \cite{Zhang2009,PhysRevLett.115.015502}.

\section{Results and discussion}

The symmetry breaking in BLG systems can be achieved by doping or applying external electric fields, which results in the appearance of a gap in the density of states. For high level doping of BLGs using boron or nitrogen, one obtains degenerate p-type or n-type semiconductors, respectively. In this case, the Fermi level is located in the conduction band or valence band, depending on the doping type. When both doping and external electric field are present, one expects a combined effect which produces either an overall enhancement or a reduction of the symmetry breaking effect. 

\begin{figure}[t]
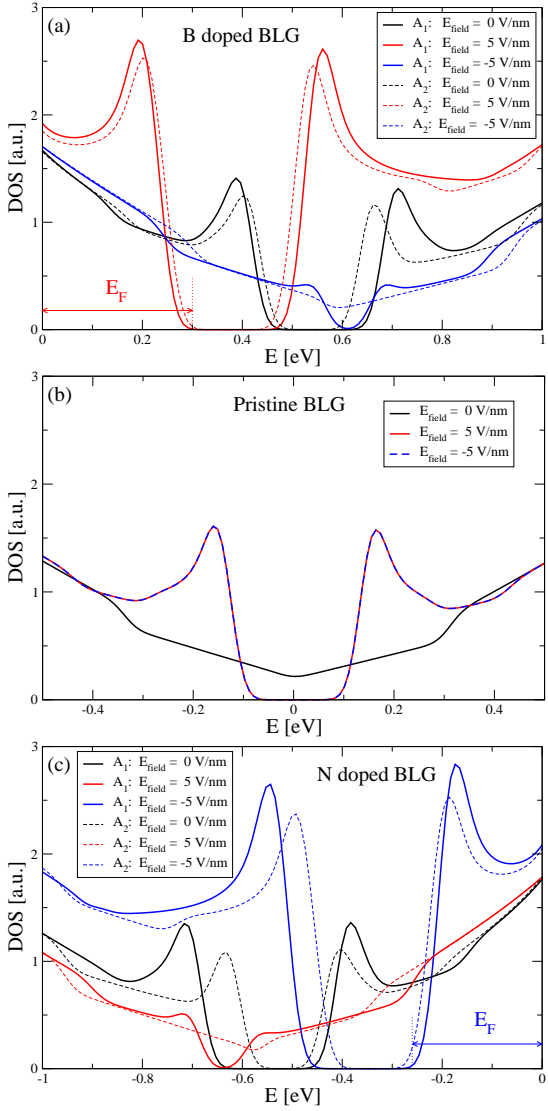

\centering
\includegraphics[width=7.1cm]{figure2a}
\includegraphics[width=7.1cm]{figure2b}
\includegraphics[width=7.1cm]{figure2c}
\caption{Density of states of BLG systems using a $5\times5$ supercell, for $E_{\rm field}= 0,\pm5$ V/nm: (a) boron doped, (b) pristine and (c) nitrogen doped. The DOS corresponding to impurities placed on $A_1$ or $B_1$ sites in the top graphene layer, are represented by solid and dashed lines, respectively. The Fermi level is taken as energy reference ($E=0$).}
\label{1xB_5x5}
\end{figure}

Figure \ref{1xB_5x5} shows the density of states for B- and N-doped BLG systems, with one impurity in the $5\times5$ supercell, as well as for the pristine BLG added for comparison. For zero electric field, the gap is absent in the case of pristine BLG, while clearly defined energy gaps are visible for the doped systems. The Fermi energy ($E_{\rm F}$) measured from the top of the valence band (B-doping) or bottom of the conduction band (N-doping) is about 0.47 eV for the $A_1$ site and slightly higher for the $B_1$ site. The energy reference ($E=0$) corresponds to the Fermi level in each case. Applying the electric field, the gaps are shifted and change their magnitude. For B-doped BLG, an electric field of $5$ V/nm, oriented along the $z$-direction, enhances the energy gap, while the Fermi energy becomes smaller, $E_F\approx0.3$ eV. By contrast, changing the orientation of the electric field to $-z$, the gap is diminished and shifted towards higher energies. Considering now the n-type doping by nitrogen atoms, the electronic properties are reversed with respect to the orientation of $E_{\rm field}$. This behavior can be correlated with the distribution of charges on the impurities and within the two graphene layers, as discussed in the following. One should also mention that boron substitutions are energetically more favorable on $B_1$ sites by 8 meV, while for nitrogen substitutions, $A_1$ sites lead to smaller energies by 19 meV.

\begin{figure}[t]
\centering
\hspace*{-6cm}(a) B-doped BLG\\
\includegraphics[height=1.3cm]{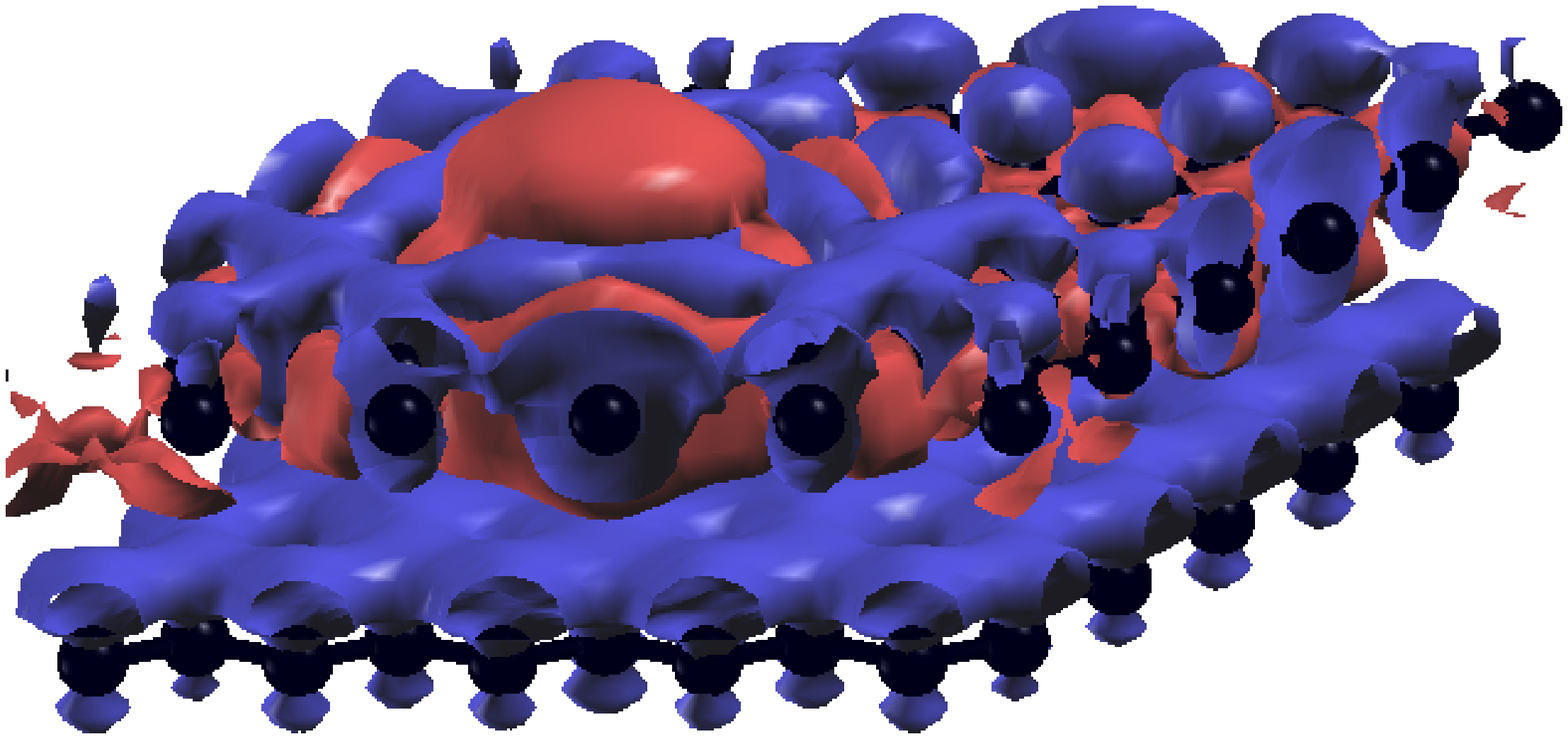}\hspace*{0.3cm}
\includegraphics[height=1.3cm]{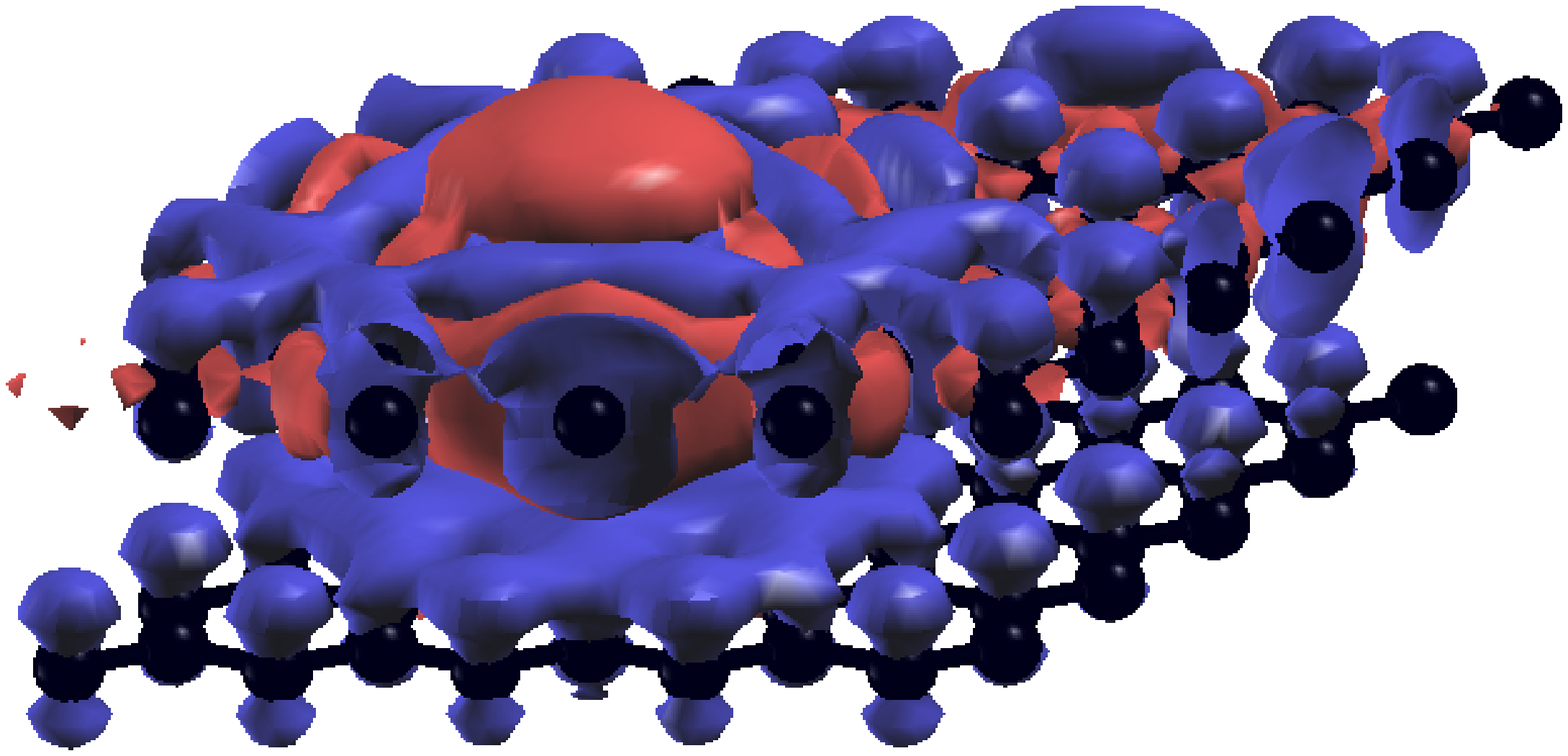}\hspace*{0.3cm}
\includegraphics[height=1.3cm]{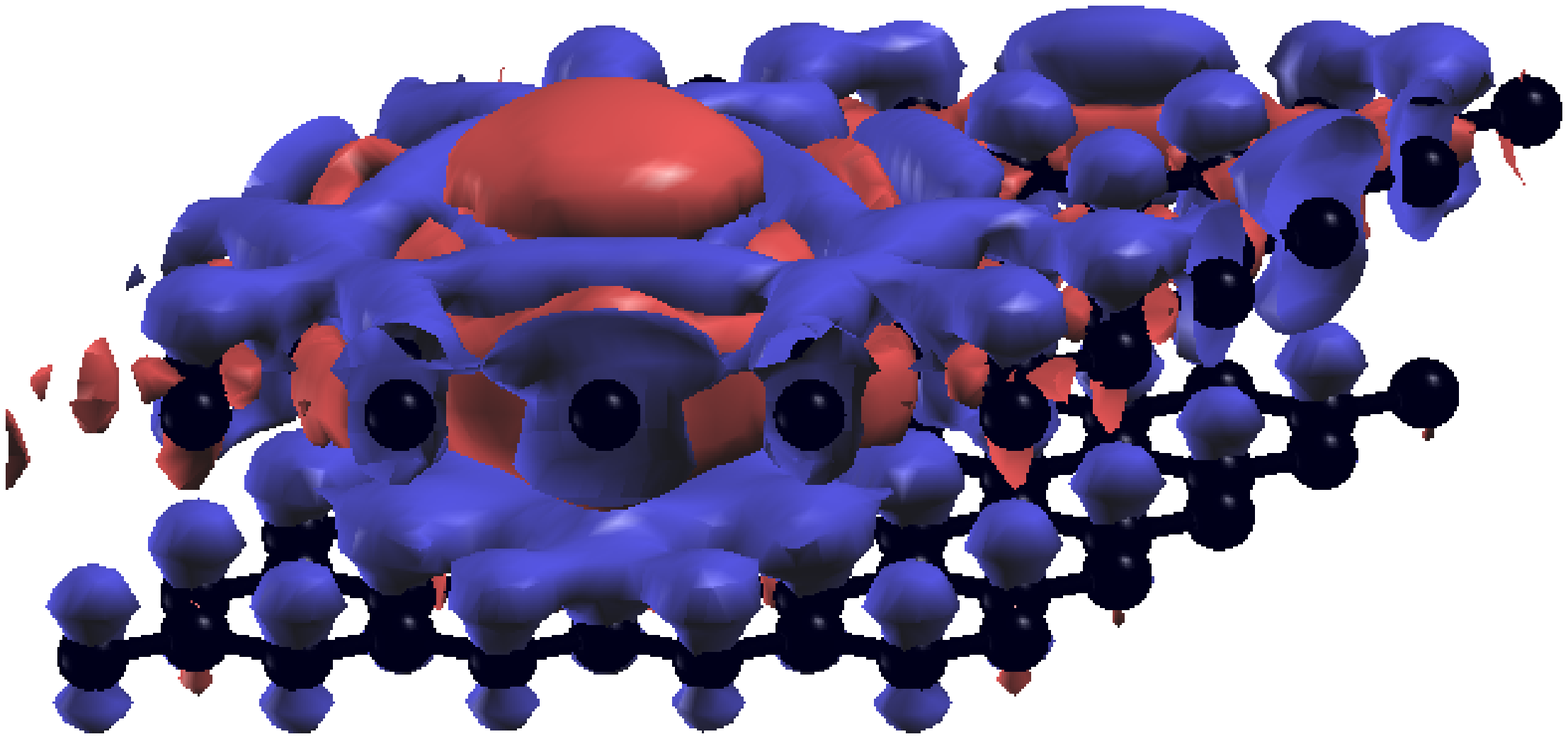}\vspace*{0.3cm}\\
\hspace*{-6cm}(b) N-doped BLG\\
\includegraphics[height=1.3cm]{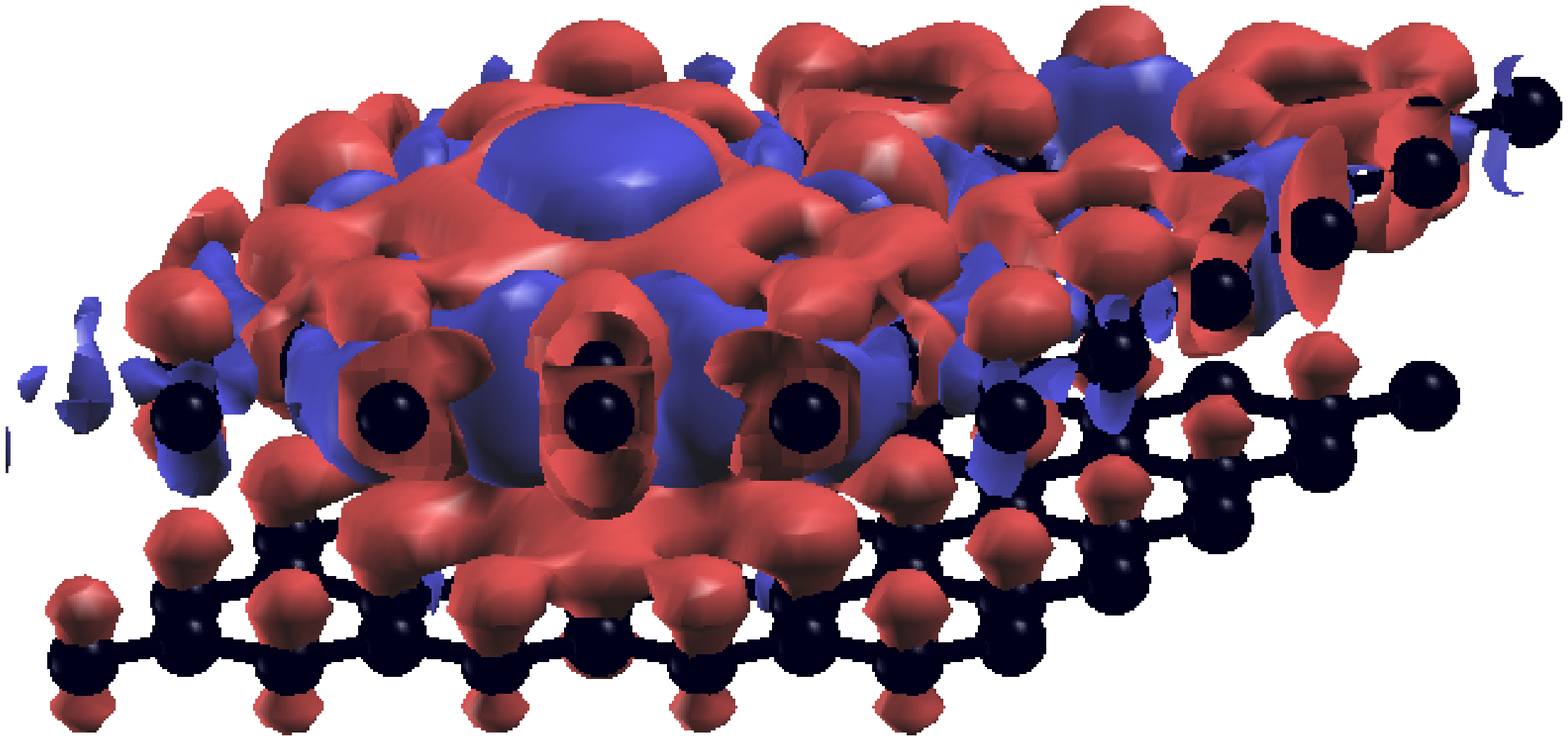}\hspace*{0.3cm}
\includegraphics[height=1.3cm]{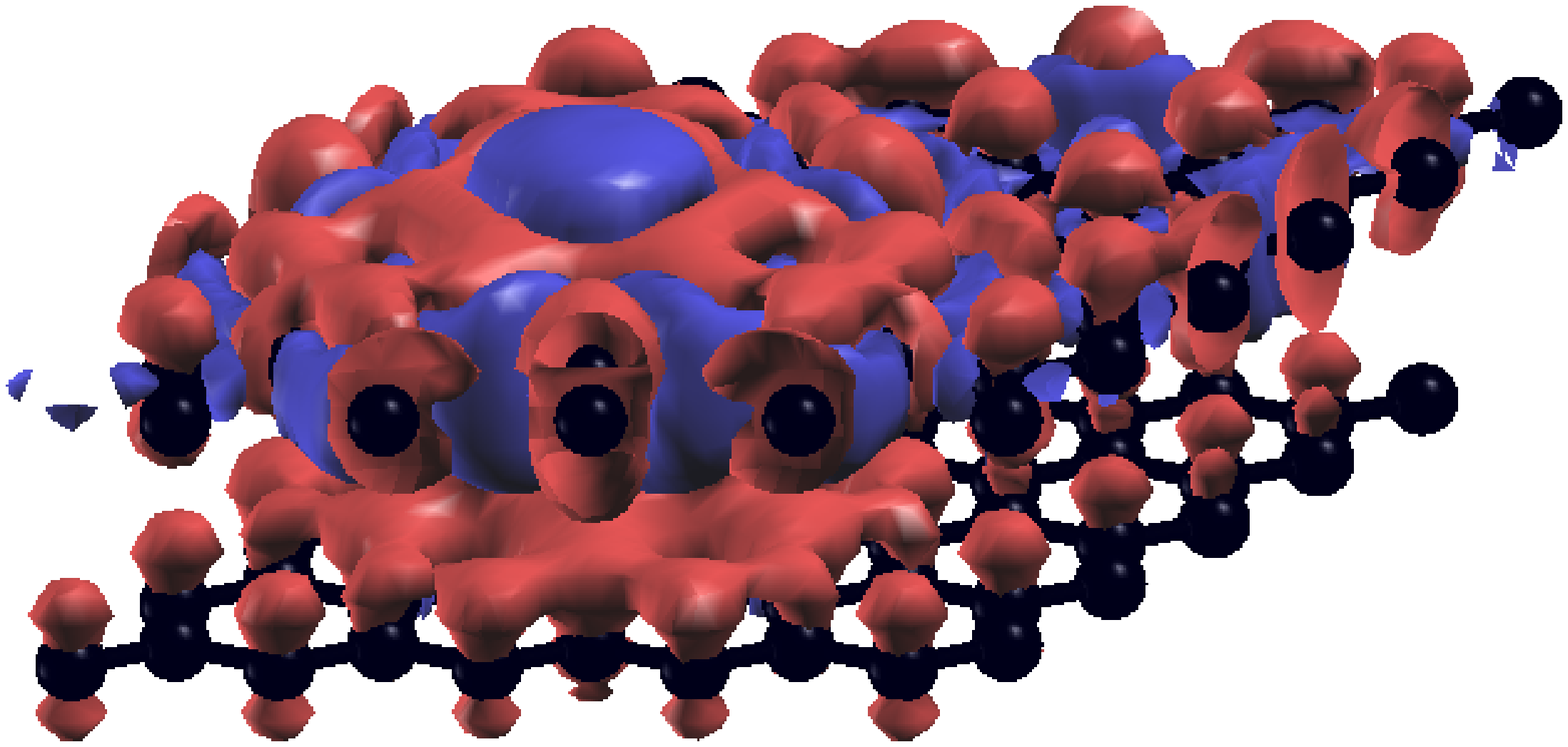}\hspace*{0.3cm}
\includegraphics[height=1.3cm]{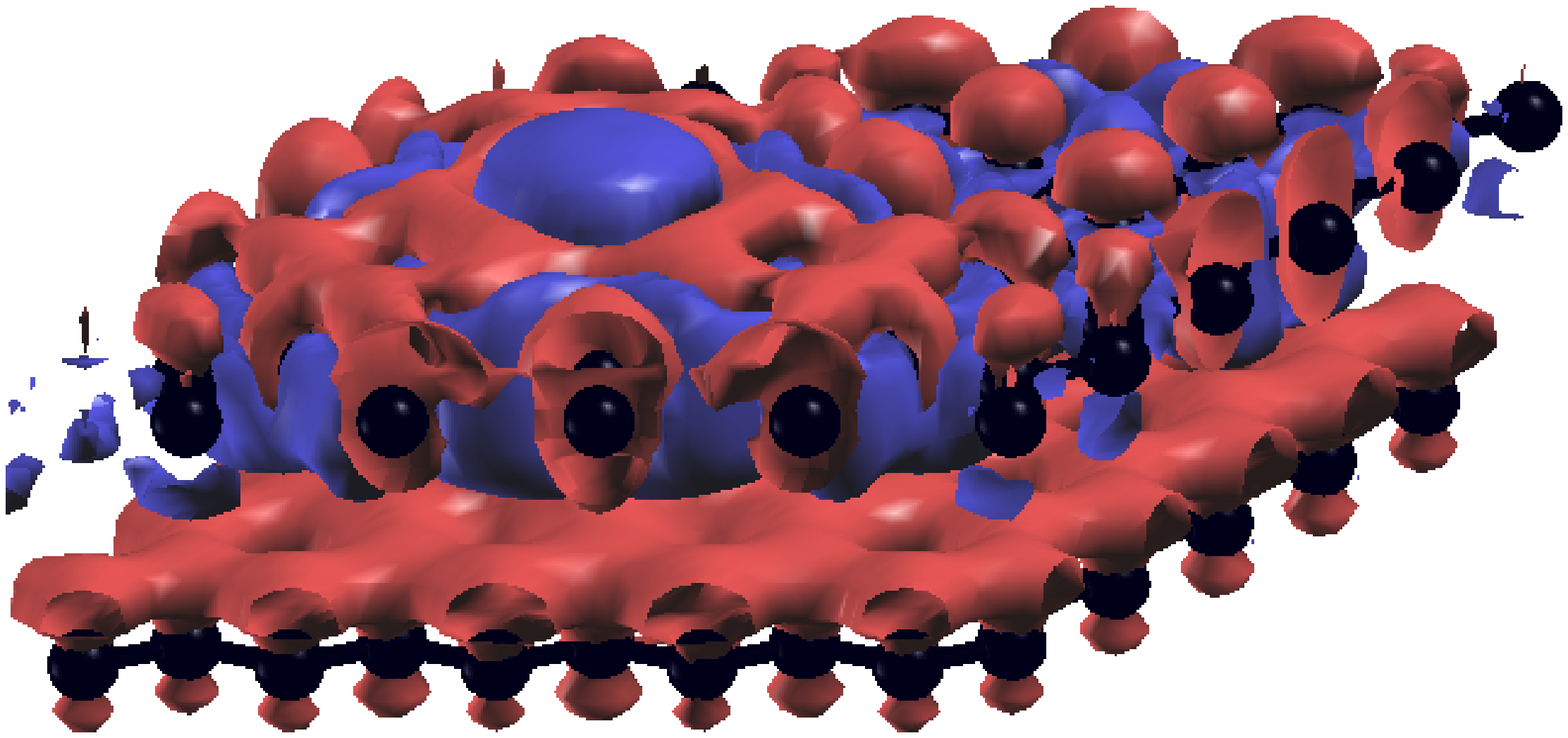}\\
\caption{Pseudocharge density maps ($\Delta\rho$) for B-doped (a) and N-doped (b) BLGs, with the extrinsic impurity present at $A_1$ site in the upper layer: from left to right the charge difference is plotted for $E_{\rm field} = -0.5, 0, 0.5$ V/nm. Significant charge variations appear in the bottom layer. The red color indicates an excess of electron charge compared to the pristine BLG, while the blue color depicts the opposite situation. The pseudocharge density differences are plotted at grid isovalue $10^{-4}$.}
\label{rho}
\end{figure}

The charge transfer between the two graphene layers in doped BLG systems is illustrated in Fig.\ \ref{rho}. The pseudocharge density difference $\Delta\rho = \rho_{B/N} - \rho_0$ describes the amount of electron charge in B- and N-doped systems ($\rho_{B/N}$) relative to the charge distribution in pristine BLG ($\rho_0$). The n- and p-type impurities concentrate a certain amount of positive and negative charge at the impurity sites, respectively. The opposite charge is found mostly within the bottom layer, just below the boron or nitrogen atoms, while alternating regions of charge surround the impurities in the top layer. This distribution of charge is visibly influenced by the presence of the electric field, specifically on its orientation, which induces an extra dipole moment. In the case of B-doped BLGs, setting $E_{\rm field}<0$ more electron charge is transferred to the upper layer, while the opposite occurs for $E_{\rm field}>0$. Therefore, in Fig.\ \ref{1xB_5x5}(a), the blue patch in the lower layer decreases as the field changes orientation, indicating that more negative charges are displaced towards the upper layer at $E_{\rm field}=-5$ V/nm compared to zero field case and the reverse trend is found for $E_{\rm field}= 5$ V/nm. Evidently, the N-doped systems show the opposite behavior in Fig.\ \ref{rho}(b).

\begin{figure}[t]
\centering
\includegraphics[width=8.5cm]{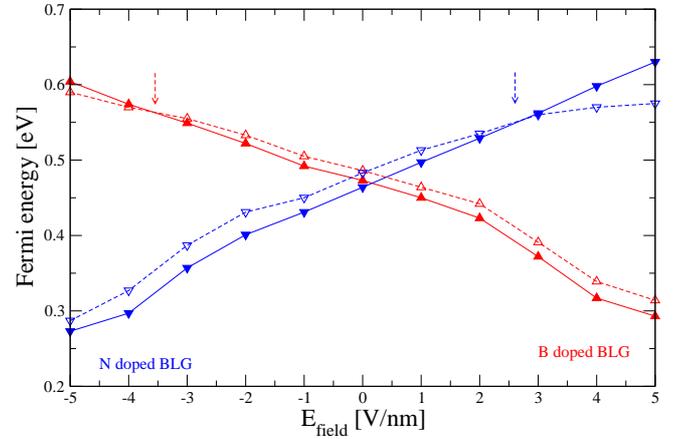}
\caption{The variation of the Fermi energy with the applied electric field, in BLG systems with one impurity in a $5\times5$ supercell. For each doping type, $A_1$ and $B_1$ sites are considered, represented by full and empty triangles, respectively. The arrows mark the closing gaps in the case of $B_1$ substitutions.}
\label{gapfield}
\end{figure}

The tuning of the Fermi energy can be achieved by gradually varying $E_{\rm field}$, as indicated in Fig.\ \ref{gapfield}. For B doping, $E_{\rm F}$ decreases as the electric field is switched from negative to positive values, in contrast to the N doping case. Basically the same trend is obtained, irrespective of the impurity site, $A_1$ or $B_1$. 
However, $B_1$ sites are less efficient in producing the interlayer asymmetry, resulting in relatively smaller gaps compared to the substitutions on $A_1$ sites (see Figs.\ \ref{EFfield-A1} and \ref{EFfield-B1}). This can be correlated with the charge distribution in the two configurations, $A_1$ and $B_1$, as described in Fig.\ \ref{rho1}, which suggests a larger dipole moment is obtained when the impurity is placed at the $A_1$ site.
 For high enough electric field magnitudes the gap is completely closed in the case of $B_1$ substitutions, which renders an enhanced metallic like behavior as the density of states becomes finite, unlike the typical degenerate semiconductor, where the Fermi energy is only about a few hundreds meV. In this case, in the absence of a gap, $E_{\rm F}$ represents the energy corresponding to the DOS minimum in absolute value, measured from the Fermi level.

\begin{figure}[h]
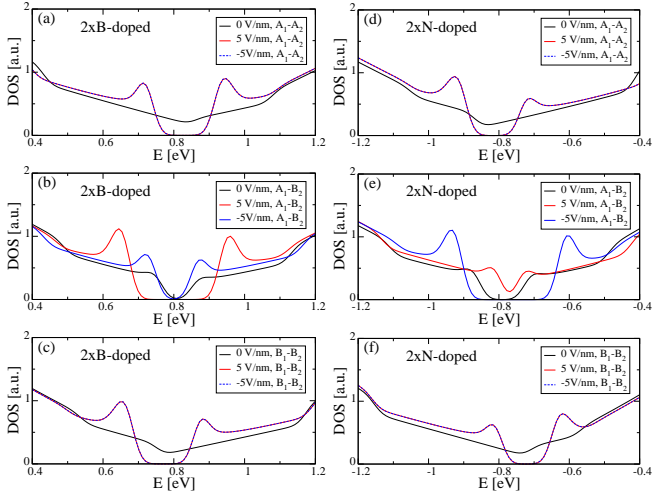

\centering
\includegraphics[width=4.2cm]{figure5a}
\includegraphics[width=4.2cm]{figure5d}\\
\includegraphics[width=4.2cm]{figure5b}
\includegraphics[width=4.2cm]{figure5e}\\
\includegraphics[width=4.2cm]{figure5c}
\includegraphics[width=4.2cm]{figure5f}\\
\caption{Density of states of BLGs with two identical impurities in the $5\times5$ supercell located at $A_1-A_2$, $A_1-B_2$ and $B_1-B_2$ sites: (a,b,c) boron pair and (d,e,f) nitrogen pair. Note that the two configurations which have the impurities in the same sublattice ($A$ or $B$) are symmetrical with respect to inter-changing the two graphene layers.}
\label{2xB_5x5}
\end{figure}

We also compare the single layer doped BLGs with systems which have substitutional boron and nitrogen impurities in both layers, placed on $A_1-A_2$, $A_1-B_2$ and $B_1-B_2$ positions. For each pair of sites the minimal distance between the two impurities was considered. If the impurities are of the same type, the largest gaps are obtained for substitutions in $A_1-B_2$ positions, as shown in Fig.\ \ref{2xB_5x5}. Compared with the previous case, where one substitutional impurity was located within the top layer, here the gap shifts are rather small, while there is still a significant variation of the gap energy with respect to the electric field magnitude and sign. However, the Fermi energy is larger, as the doping concentration doubles. The asymmetry introduced in the $A_1-B_2$ configuration introduces a small gap in the absence of an electric field, in contrast to $A_1-A_2$ and $B_1-B_2$ configurations, where the DOS is quite similar with the one of pristine graphene, i.e. gap-less but shifted with respect to $E_F$. One should also note that for $A_1-A_2$ and $B_1-B_2$ configuration the two orientations of the electric field are equivalent. Regarding the formation energies, similarly to the systems with one impurity in the supercell, for boron substitutions the $B_1-B_2$ configuration yields the lowest energy, closely followed by $A_1-B_2$ (+5 meV) and $A_1-A_2$ (+6 meV). For nitrogen substitutions, the sequence of total energies is reversed: taking $A_1-A_2$ configuration as reference, we find higher energies for $A_1-B_2$ (+22 meV) and $B_1-B_2$ (+41 meV). 

If the two impurities are different, e.g. boron in the top layer and nitrogen in the bottom layer, the BLG system becomes an intrinsic semiconductor, with a small gap ($\sim 0.1-0.2$ eV) at zero electric field as depicted in 
Fig.\ \ref{BN}, with a small variation due to doping positions. In this case, the B-N pair induces a local field and it was found that $A_1-A_2$ substitutions are most effective in inducing the energy gap, compared to the other two. Applying the external electric field, there is a relatively small gap tuning, although the density of states changes significantly. Concerning the different types of substitutions the same trend is obtained independent on $E_{\rm field}$. One expects that additional boron and nitrogen substitutions, in equal amounts, would further enhance the semiconducting gap of the modified BLG. If the extrinsic impurities are present in the same layer the formation of embedded hexagonal boron nitride domains may further enhance the gap.  

\begin{figure}[t]
\centering
\includegraphics[width=8.5cm]{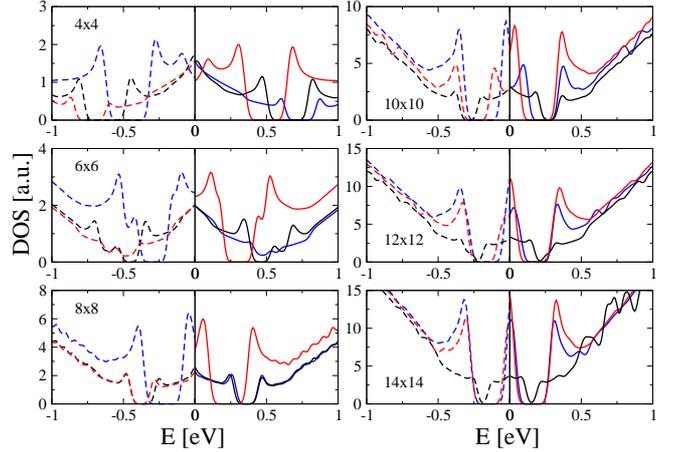}
\caption{The effect of the doping concentration under external electric field of 5 V/nm (red), 0 V/nm (black) and -5 V/nm (blue), for B-doping (solid) and N-doping (dashed) BLGs. The extrinsic impurity was placed in the top layer, in $A_1$ position, in supercells ranging from $4\times4$ to $14\times14$. The gaps are gradually shifting towards the Fermi level, while the field orientation becomes less important.}
\label{dc}
\end{figure}

\begin{figure}[t]
\centering
\includegraphics[width=8.5cm]{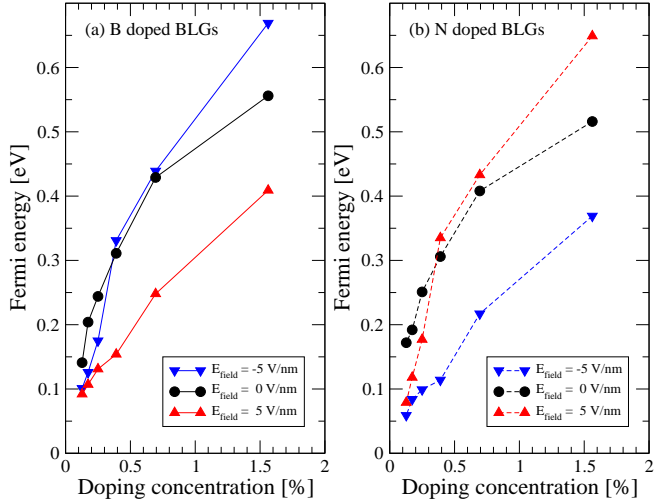}
\caption{The variation of the Fermi energy with the doping concentration, for (a) boron and (b) nitrogen doping, with and without the applied electric field. The $E_{\rm F}$ values correspond to the structures discussed in Fig.\ \ref{dc}.}
\label{dopingconc}
\end{figure}

An important technological aspect concerns the field effect in the context of varying the doping concentration. One question that arises is to what extent doped BLGs retain the conductive behavior under applied electric field. By increasing the supercell dimension, while keeping one impurity in the top layer, we reduce the doping towards the intrinsic limit. Structures with substitutional impurities placed on $A_1$ positions are investigated. Figure \ref{dc} shows the DOS of B- and N-doped BLG systems with unit cells ranging from $4\times4$ to $14\times14$, which corresponds to lowering the doping concentration from 1:64 to 1:784. In the high doping limit, as previously discussed, wide gaps are visible even in the absence of external field, while for $E_{\rm field}= \pm 5 $ V/nm the gaps are shifted in opposite directions. The DOS near the gap edges is enhanced for one field orientation and diminished for the other one. As the doping concentration decreases, the Fermi energy is getting smaller, from $\sim500$ meV ($4\times4$) to less than $\sim50$ meV ($14\times14$). 
At very low doping concentrations, the gaps tend to vanish when no electric field is applied. In addition, the orientation of the field becomes less and less relevant, as the gap shift decreases. At this point the effects produced by the external electric field are beginning to dominate the local fields due to the extrinsic impurities, as one can see from Fig.\ \ref{dopingconc}.  
If one further lowers the doping concentration, it is expected that for a finite electric field, the Fermi level will enter the gap, resembling the intrinsic BLG with non-degenerate semiconductor behavior. Our results suggest that this transition may be indeed achieved provided the doping concentration is sufficiently low and the electric field high enough.
At the same doping concentration in BLG systems with nitrogen substitutions the Fermi energy is comparatively smaller. One should also note from Fig.\ \ref{dopingconc} that for a doping concentration of $\sim$0.35 \%, for both types of impurities, the Fermi energy at an applied electric field of $\pm$5 V/nm is the same as without field. This point marks the doping concentration value below which the action of the electric field overcomes the effects introduced by impurities. 

In order to analyze comparatively the stability of the boron and nitrogen doped systems, with concentrations between 0.13\% ($14\times14$) and 1.56\% ($4\times4$), we calculate the formation energies, defined as $E_{\rm f} = E_{\rm d-BLG} - \sum_i n_{\rm X_i} \mu_{\rm X_i}$, with ${\rm X_i}$ = C, B or N, where $E_{\rm d-BLG}$ is the total energy of the doped BLG and $\mu_{\rm C}$, $\mu_{\rm B}$, $\mu_{\rm N}$ are the chemical potentials in pristine BLG, $\alpha$-boron crystal and $N_2$ molecule, respectively. For the largest doping concentration, we find for boron substitution in the $4\times4$ system a formation energy of 1.02 eV, while for nitrogen substitutions a significantly smaller formation energy of 0.44 eV is obtained. These values are close to the ones reported for boron doped graphene monolayer systems of the same supercell size \cite{doi:10.1021/acs.jpcc.5b07359,doi:10.1021.jp106764h,doi:10.1063/1.5018065} and nitrogen doped graphene \cite{doi:10.1021/acs.jpcc.5b07359,Lv2012}, indicating that N substitutions are more stable. However it should be noted that by decreasing the doping concentration, in the dilute limit corresponding to the largest system ($14\times14$), $E_{\rm f}$ is reduced by $\sim$0.3 eV for both boron and nitrogen doping, a trend which was also observed in B/N doped graphene monolayer systems \cite{doi:10.1021/acs.jpcc.5b07359,doi:10.1021.jp106764h}.

A few comments should be made regarding potential disorder effects. Compared to configuration averages performed by extensive {\it ab initio} calculations using large supercells, a tight-binding model calibrated on DFT based calculations represents a more feasible approach \cite{PhysRevB.82.245414}. Here it was shown that if random disorder is applied, by distributing boron atoms in one of the layers, the gap still persists and it is shifted towards higher energies. This analysis was performed for the zero field case, revealing also the importance of impurity distribution amongst the two sublattices. In connection to potential variations of the electronic properties related to the distribution amongst the two sites, we investigate in more detail a $10\times10$ system, doubling the linear size of a $5\times5$ system. Different configurations of impurities, from {\it all}-$B_1$ positions to {\it all}-$A_1$ positions, labeled as A and E configurations, respectively, are considered as depicted in Fig.\ \ref{disorder}. We find that the intermediate configurations, denoted by B, C and D, exhibit a rather similar behavior to the completely ordered instances, except for some minor changes in the Fermi energy and gap magnitude (Fig.\ \ref{DOSdisorder}). Therefore, we conclude that, overall, the substitutional sites are less relevant to the electronic properties of doped BLGs, at least in the limit of weak disorder. More importantly, the energy gap can be tuned by both doping concentration and external field, which both parametrize the transition between the semiconducting and the metallic like behavior.

\section{Conclusions}

We investigated the combined effect of doping and external electric fields in BLG systems. The extrinsic impurities produce the symmetry breaking between the two graphene layers, inducing gaps in the valence or the conduction band, which can be further controlled by the external electric field. Highly doped BLGs with boron and nitrogen behave as degenerate semiconductors with a tunable Fermi energy, while by lowering the doping concentration the non-degenerate behavior can be recovered, particularly when a proper electric field is applied. For low enough impurity doping or for preferential sublattice substitutions present in both graphene layers the gap may close in the absence of the electric field. In this context, a semiconductor - metal like transition in doped BLGs may be triggered by the external electric field, depending also on the extrinsic doping. Our results suggest the possibility of adjusting the effective doping in BLGs by employing electric fields, which has a twofold implication in potential BLG based integrated circuits. First, since the p-n, p-i-n junctions and the contact lines using degenerate semiconductors are primary structures in the standard CMOS technology, achieving reconfigurable device properties of basic electronic components is an important goal which shall be advocated for also in a bi- or multi-layer graphene based technology. Secondly, by using stacked gates one can influence the conductive behavior even in highly doped regions, by externally setting the magnitude of the electric field, while fixing the doping concentration.  \\

{\bf Acknowledgments} \\

This work was supported by the Romanian Ministry of Research and Innovation under the project PN 18090205/2018 and by Romania-JINR cooperation project JINR Order 322/21.05.2018, no. 29.



\bibliography{manuscript_R1}

\newpage

\onecolumn


\appendix
\section{Supplementary Material}

\begin{figure}[h]
\centering
\includegraphics[width=11cm]{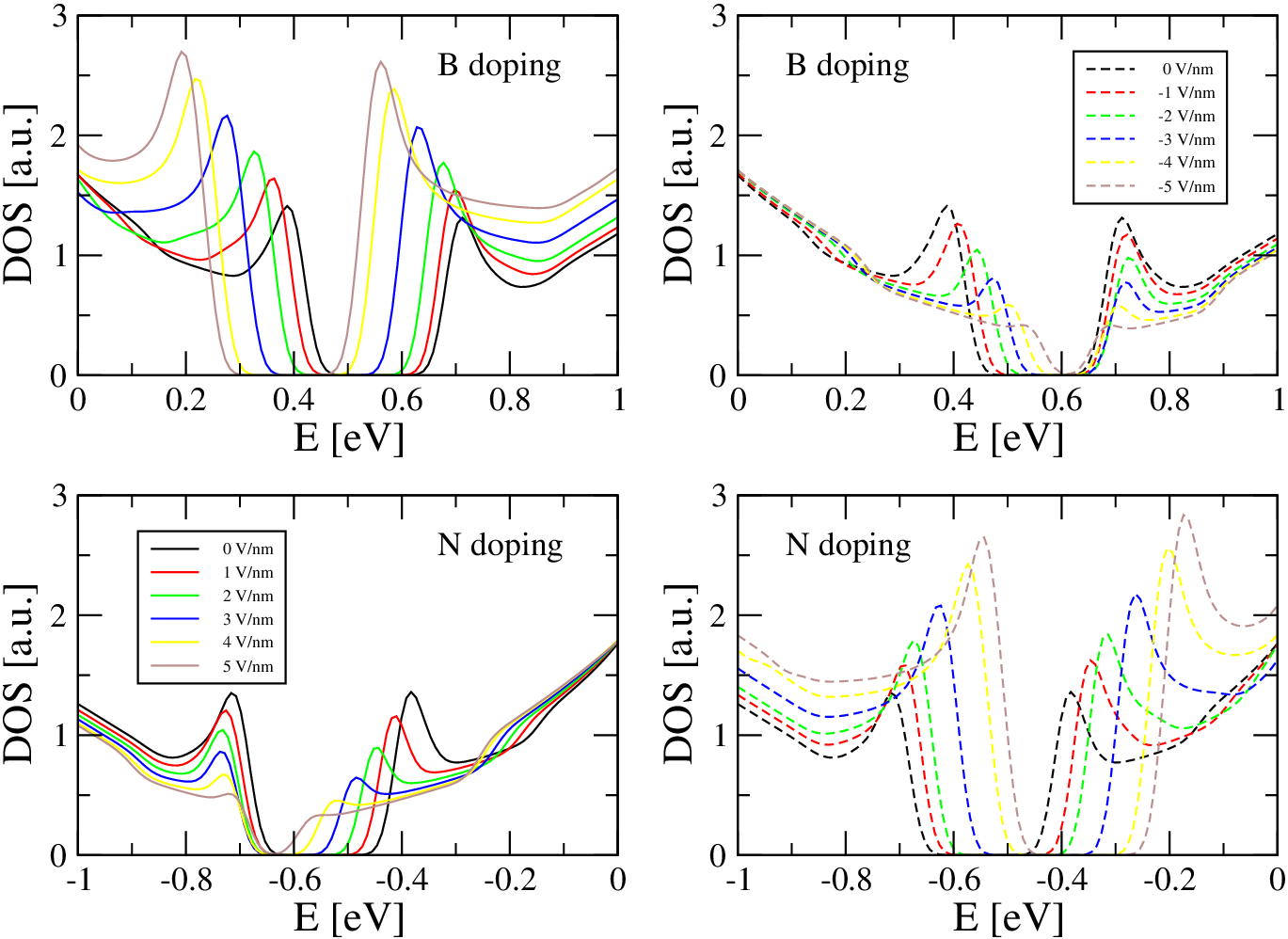}
\caption{Density of states of BLG with one impurity in the $5\times5$ supercell located at $A_1$ site. The Fermi level corresponds to $E=0$.}
\label{EFfield-A1}
\end{figure}

\begin{figure}[h]
\centering
\includegraphics[width=11cm]{figure_S2}
\caption{Density of states of BLG with one impurity in the $5\times5$ supercell located at $B_1$ site. The Fermi level corresponds to $E=0$.}
\label{EFfield-B1}
\end{figure}

\begin{figure}[t]
\centering
\includegraphics[width=7.5cm]{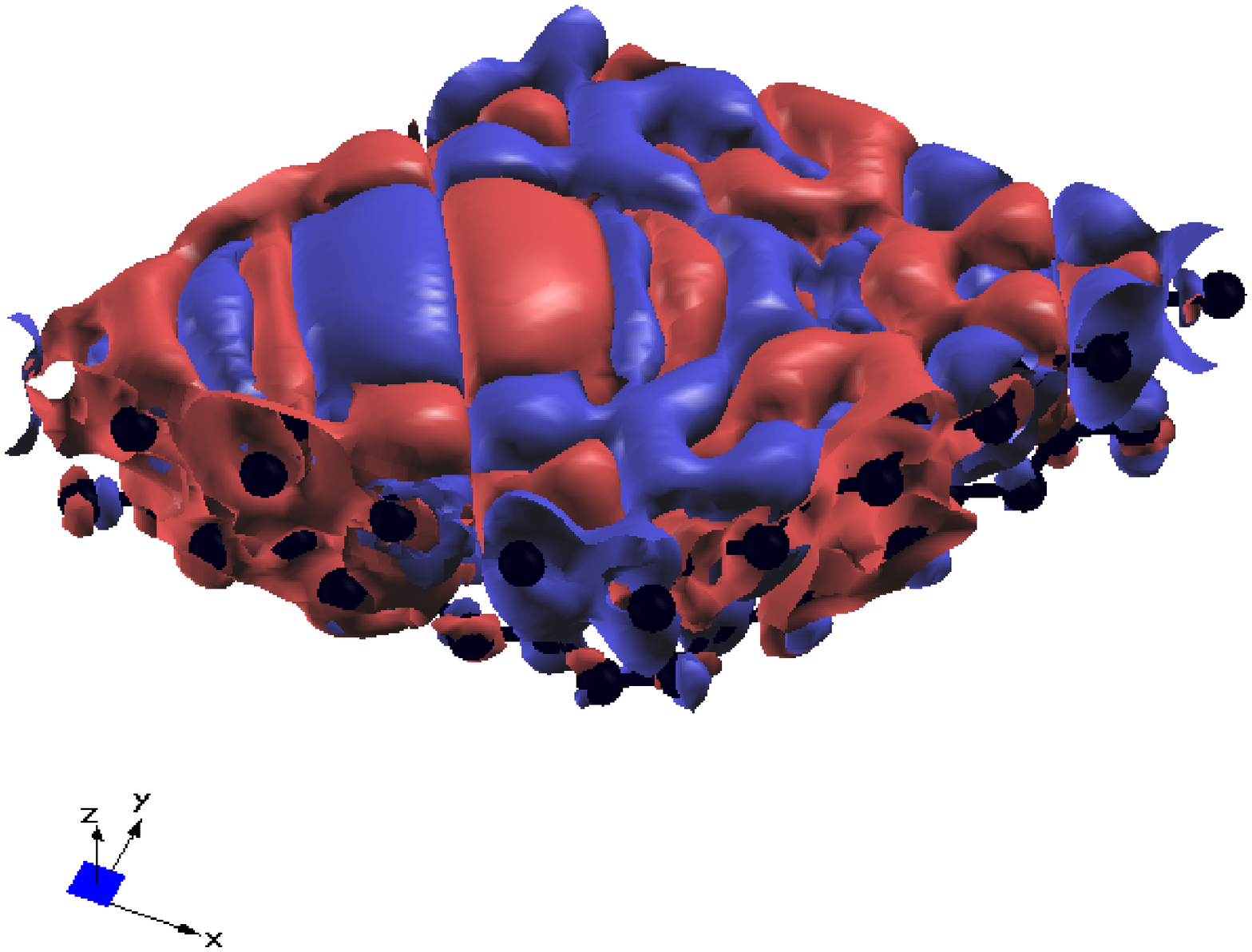}
\includegraphics[width=7.5cm]{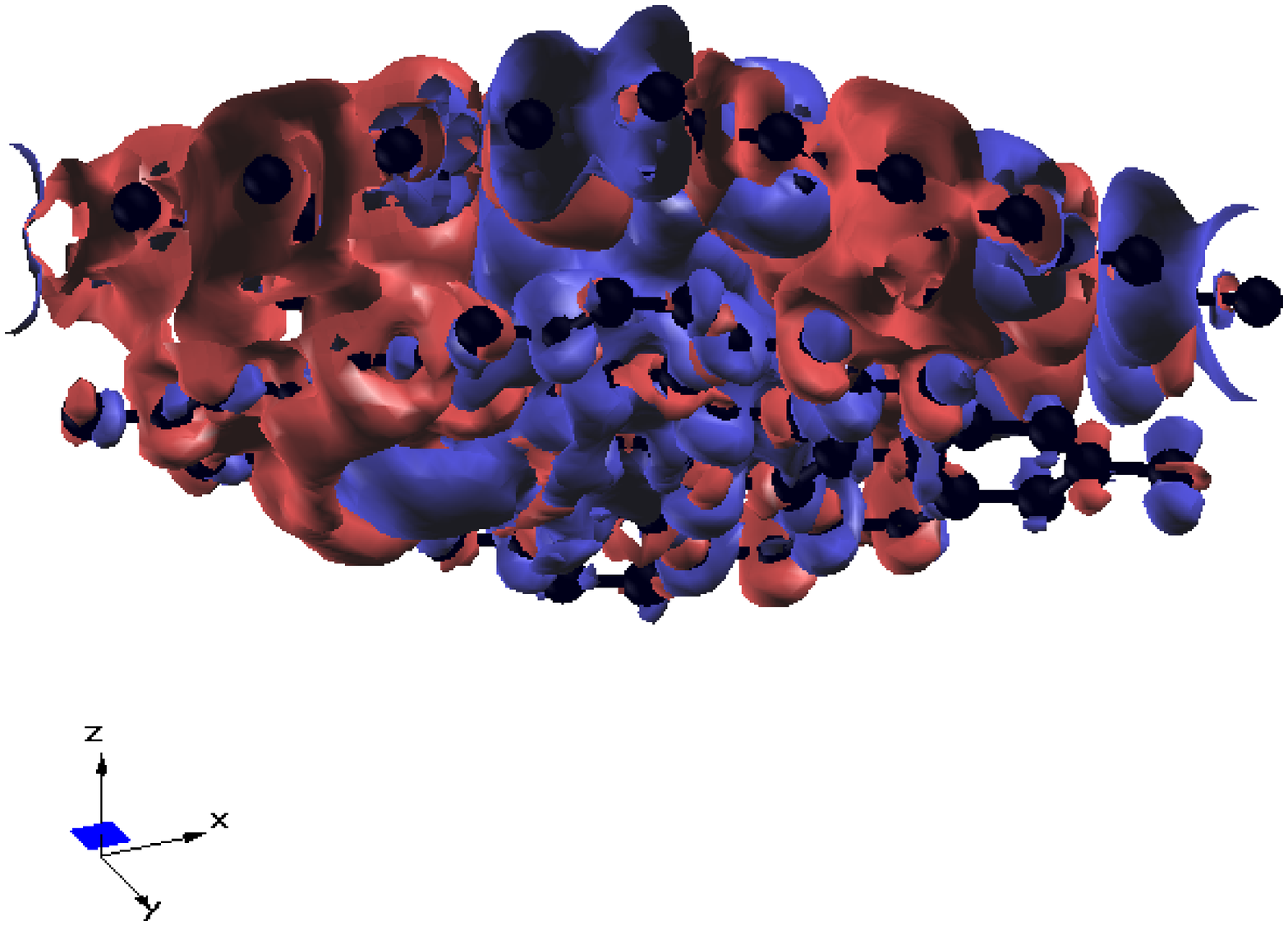}\\
(a)\hspace*{7cm}(b) \vspace*{0.5cm}\\
\includegraphics[width=7.5cm]{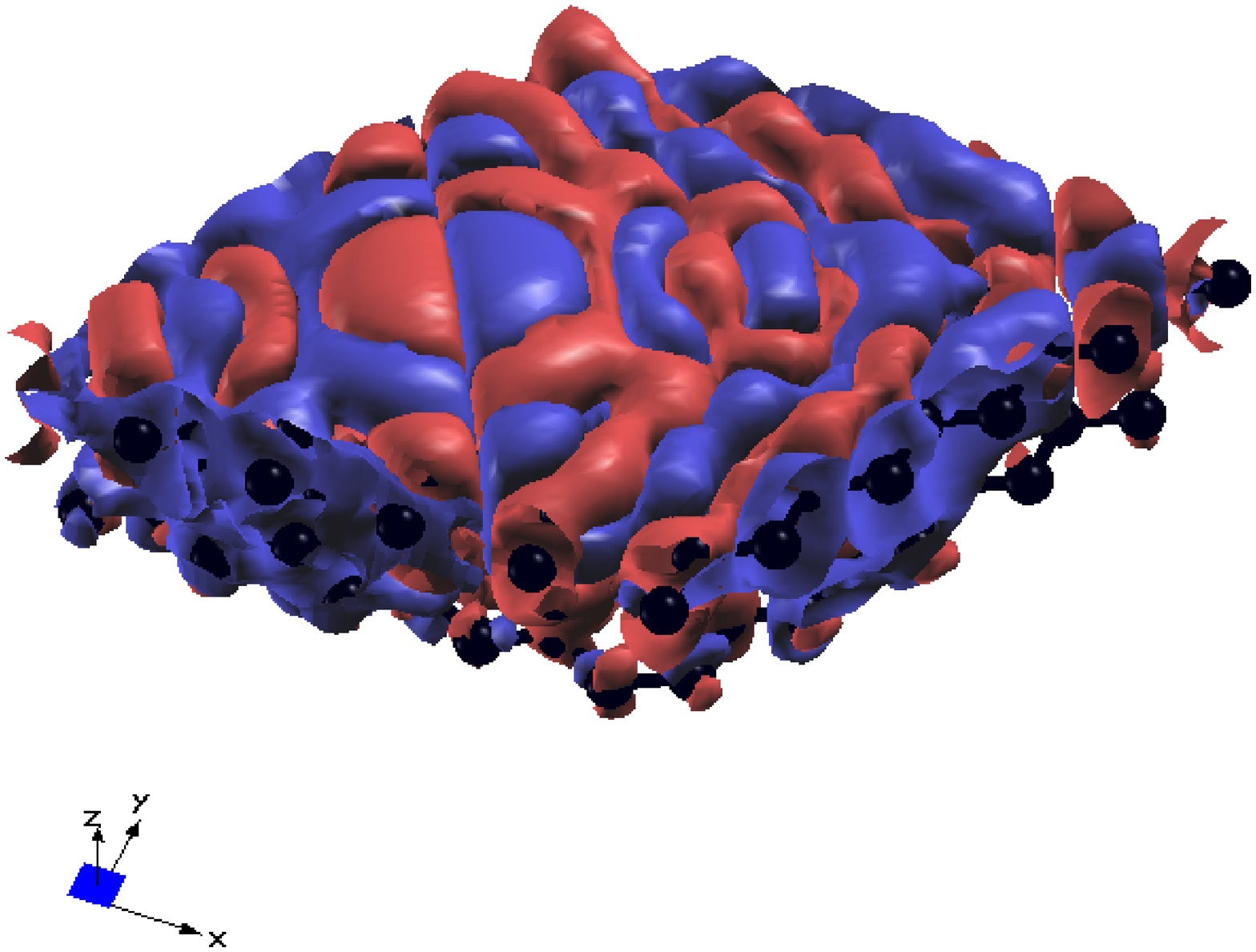}
\includegraphics[width=7.5cm]{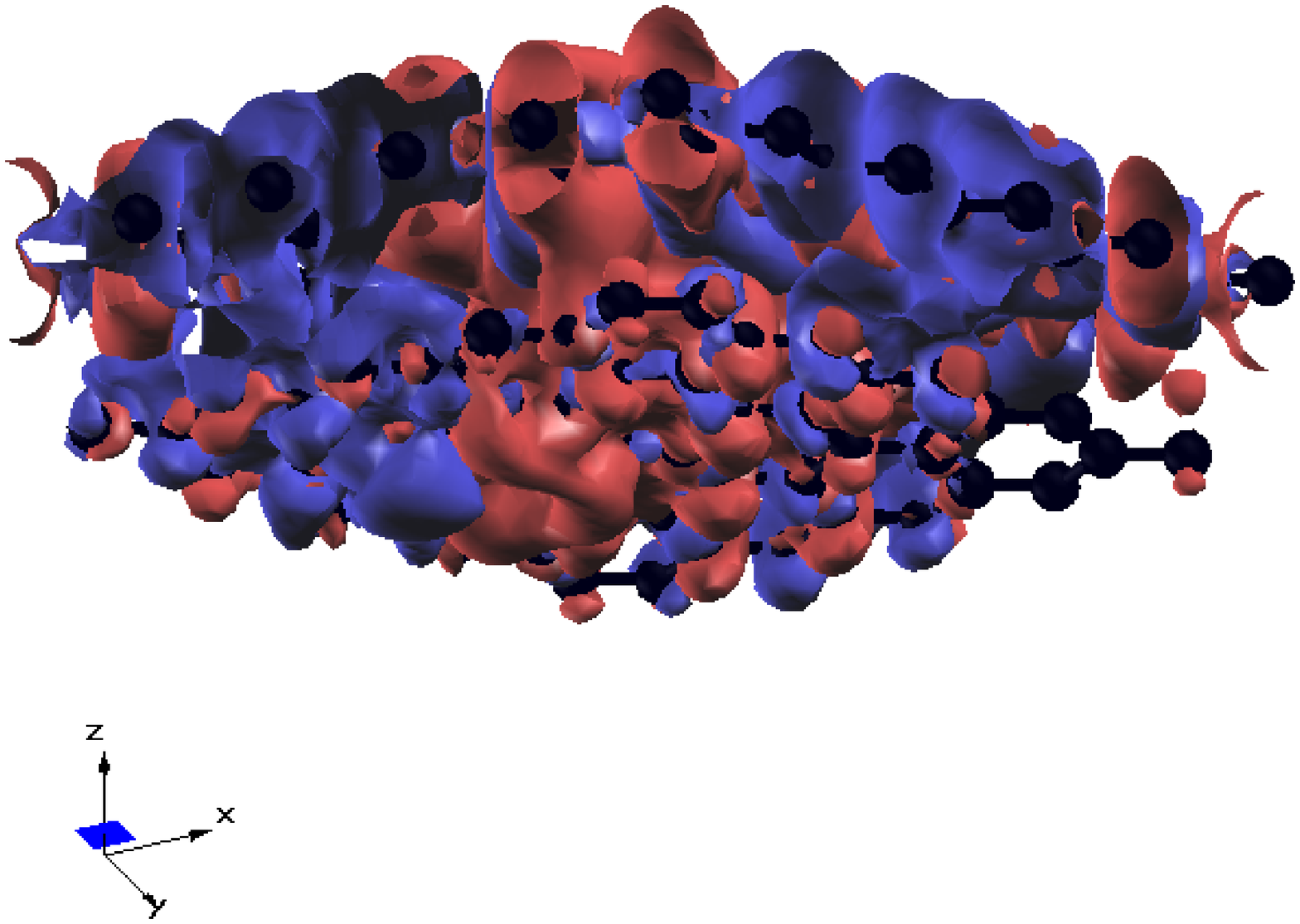}\\
(c)\hspace*{7cm}(d) \vspace*{0.5cm}\\
\caption{Pseudocharge density difference ($\Delta\rho_{AB}$) between two $5\times5$ systems, one with a B/N impurity at $A_1$ site and one with the same impurity placed at $B_1$ site, i.e. $\Delta\rho_{AB}=\rho_{B/N}(A_1) - \rho_{B/N}(B_1)$, for $E_{\rm field}=0$ V/nm, plotted at grid isovalue of $10^{-5}$: (a,b) boron and (c,d) nitrogen impurities. Top layer view (left) and bottom layer view (right) are presented. For the $A_1$ type system, in the case of boron substitution, there is less electronic charge localized on the bottom layer compared to the $A_2$ type system. More positive charge in the bottom layer in the context of negatively charged p-type impurity in the top layer induces a larger asymmetry between the two layers in the $A_1$ configuration. The opposite is found for the nitrogen doped system.}
\label{rho1}
\end{figure}

\begin{figure}[t]
\centering
\includegraphics[width=14cm]{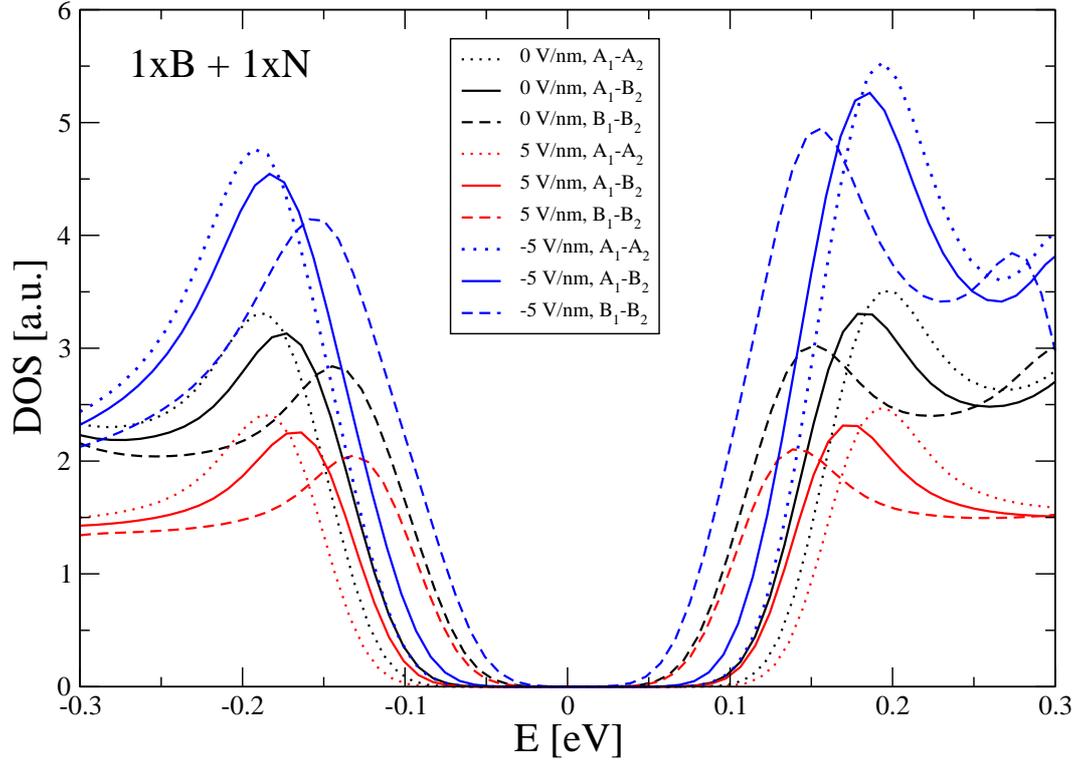}
\caption{Density of states of BLGs with two impurities in the $5\times5$ supercell: one B impurity in the upper layer and one N impurity in the lower layer, for $A_1-A_2$ (dotted), $A_1-B_2$ (solid) and $B_1-B_2$ (dashed) configurations.}
\label{BN}
\end{figure}


\begin{figure}[t]
\centering
\includegraphics[width=5cm]{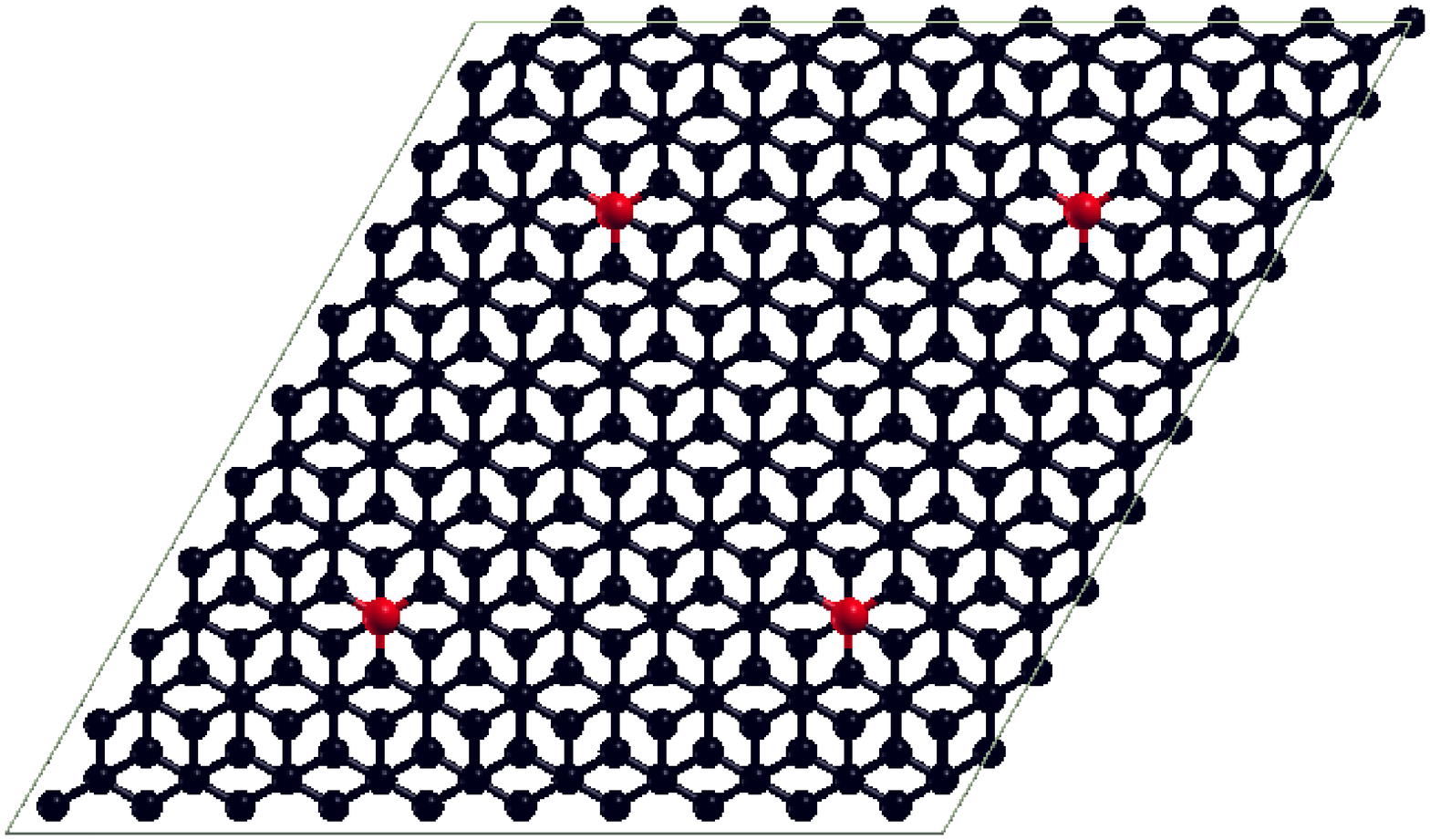}
\includegraphics[width=5cm]{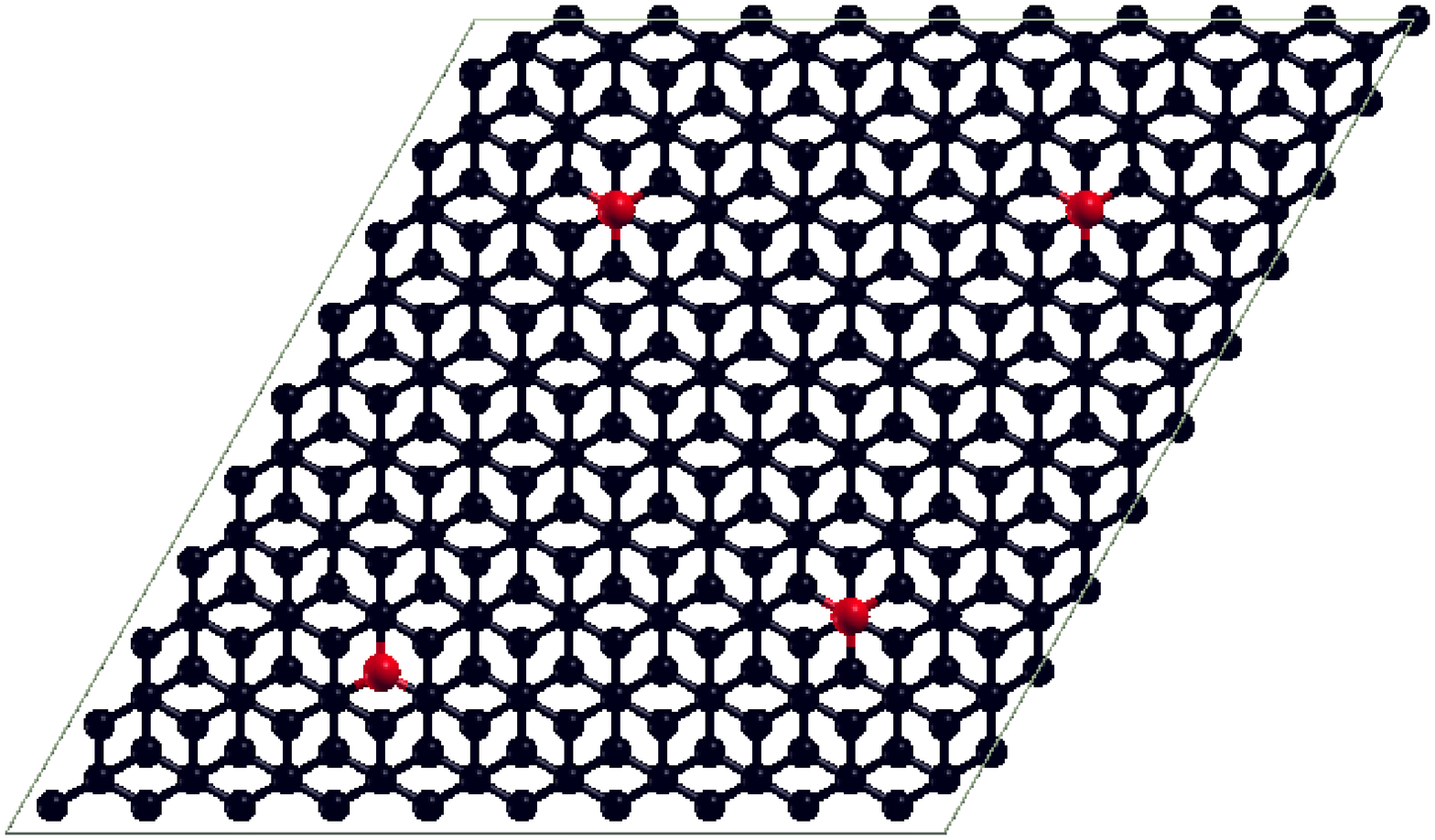}
\includegraphics[width=5cm]{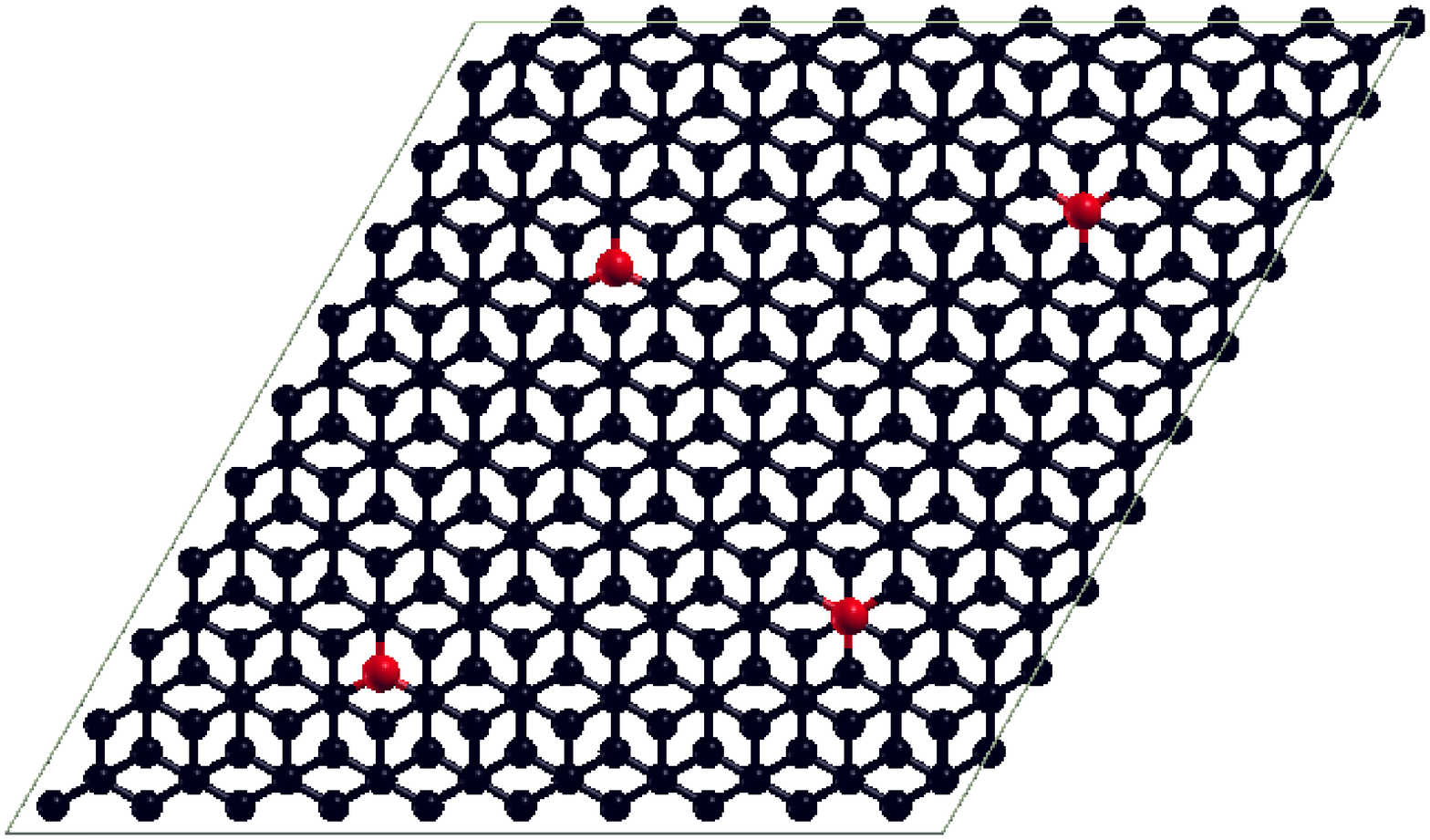}\\
\hspace*{-1.5cm}A\hspace*{5cm}B\hspace*{5cm}C \vspace*{0.3cm}\\
\includegraphics[width=5cm]{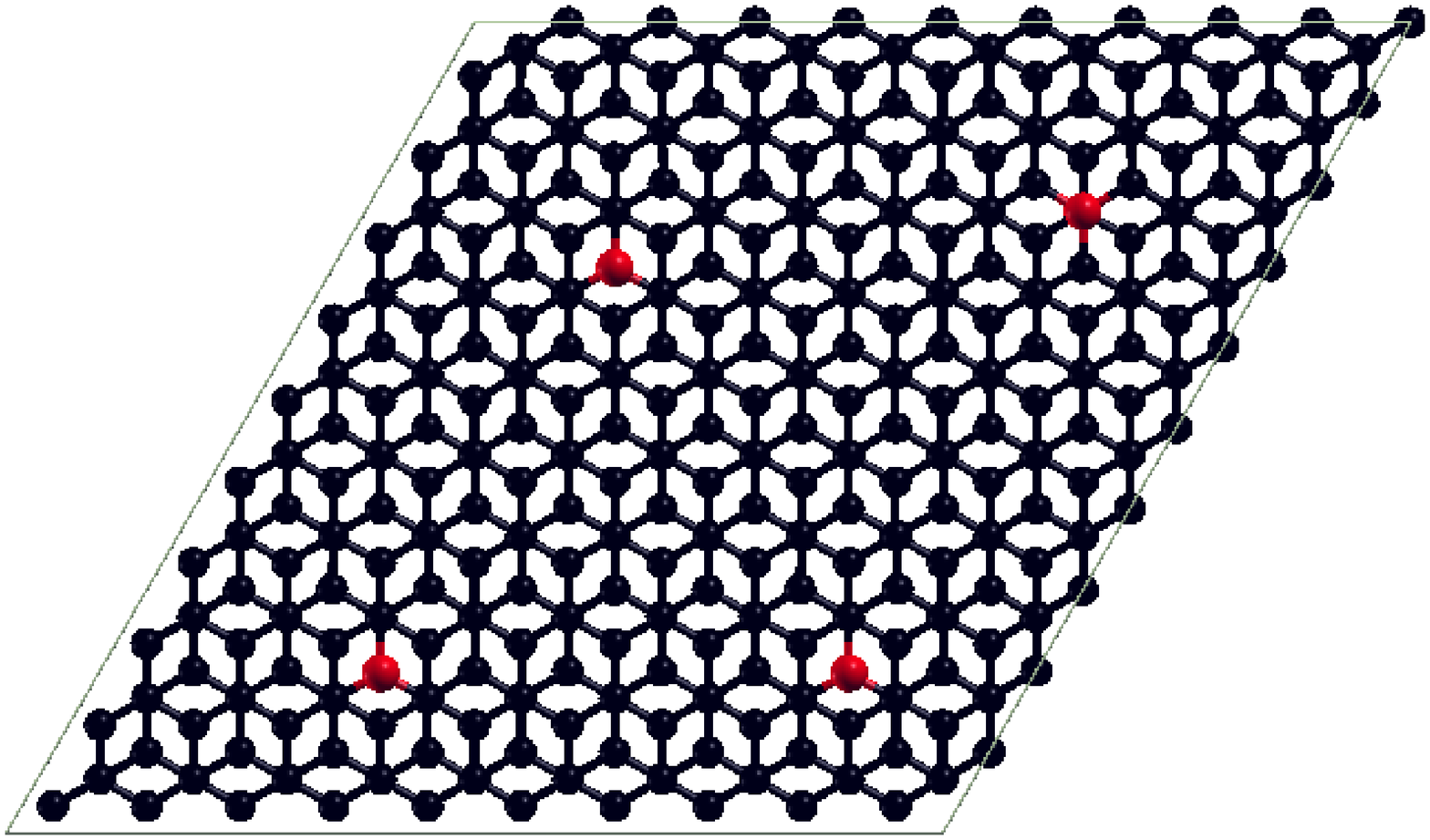}
\includegraphics[width=5cm]{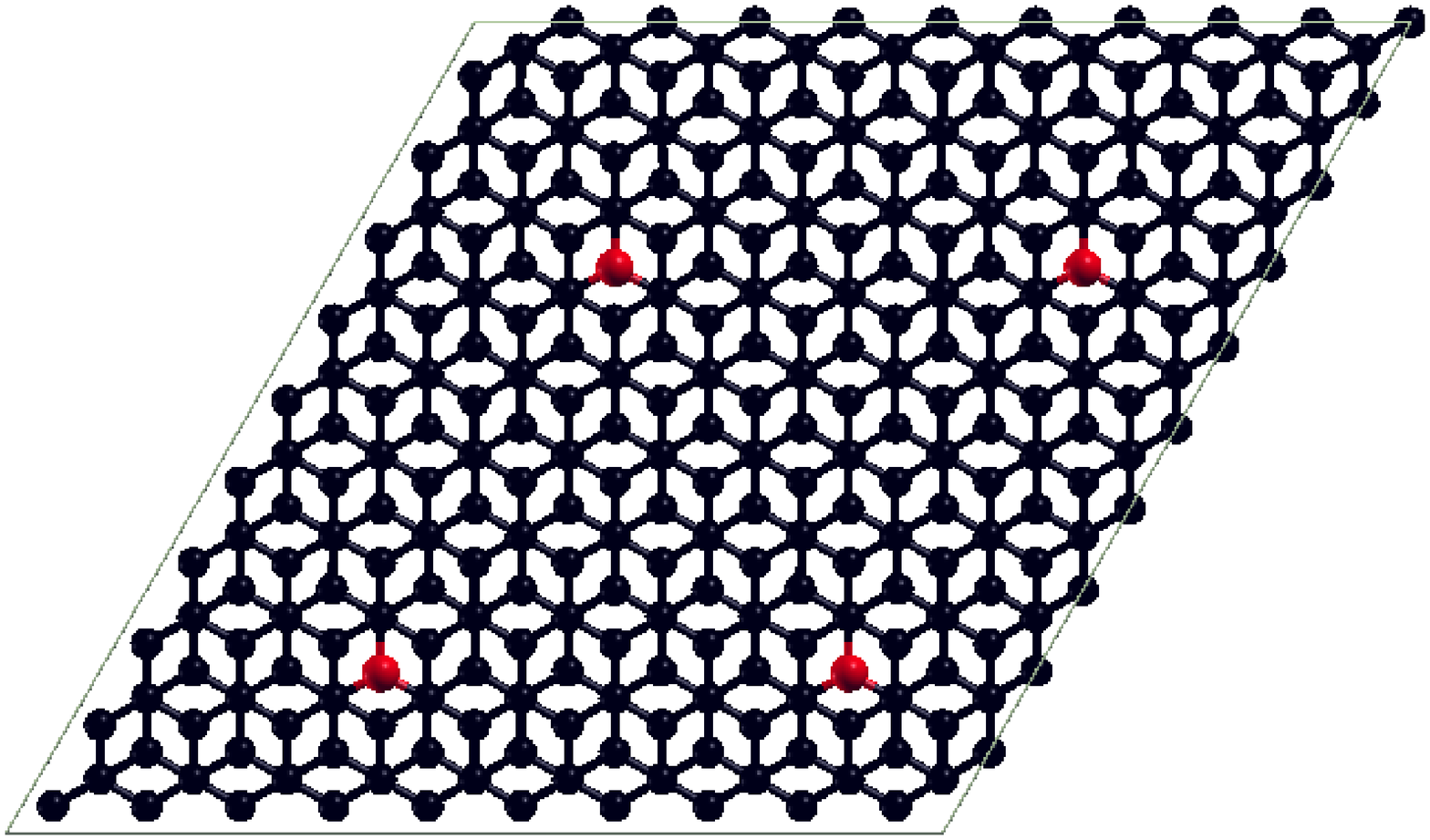}\\
\hspace*{-1.5cm}D\hspace*{5cm}E
\caption{BLGs doped with boron, with 4 impurities in a $10\times10$ supercell, in five configurations (A-E). Configuration A, with four boron atoms on $B_1$, is gradually changed to configuration E, with impurity atoms placed in $A_1$ positions. A and E configurations are equivalent to the $5\times5$ systems with one impurity placed at $B_1$ and $A_1$ sites, respectively.}
\label{disorder}
\end{figure}

\begin{figure}[t]
\centering
\includegraphics[width=12cm]{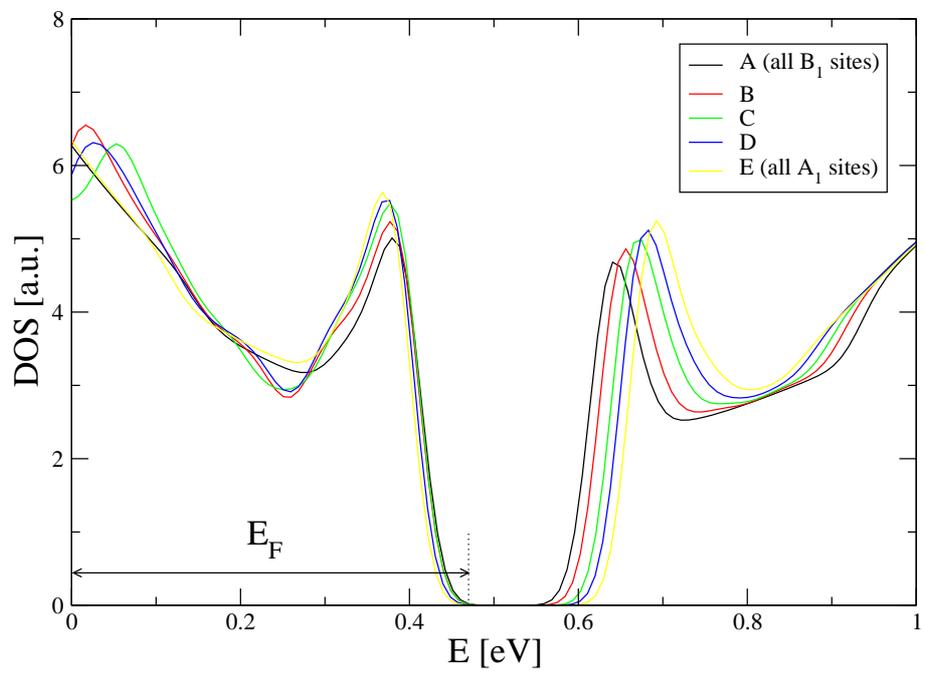}
\caption{Density of states of BLGs shown in Fig.\ \ref{disorder}. The gap increases in the A-E sequence of configurations, while the Fermi energy remains practically the same ($\sim 0.47$ eV). }
\label{DOSdisorder}
\end{figure}


\end{document}